\documentclass[11pt,twoside]{article}
\usepackage[pdftex]{graphicx}
\usepackage{amsmath}
\usepackage{amssymb}
\usepackage{cite}
\usepackage{mathrsfs}

\begin{document}

\thispagestyle{plain}

\begin{center} {\Large \bf
\begin{tabular}{c}
Optical tomography of Fock state superpositions \\[-1mm]
\end{tabular}
 } \end{center}

\smallskip

\begin{center} {\bf Sergey N. Filippov and Vladimir I. Man'ko}
\end{center}

\smallskip

\begin{center}
{\it
Moscow Institute of Physics and Technology (State University)\\
Institutskii per. 9, Dolgoprudnyi, Moscow Region 141700, Russia\\
{\rm and}\\
P.~N.~Lebedev Physical Institute, Russian Academy of Sciences\\
Leninskii Prospect 53, Moscow 119991, Russia}

\smallskip

e-mail:~~~sergey.filippov~@~phystech.edu,~~~manko~@~sci.lebedev.ru
\end{center}

\begin{abstract}\noindent
We consider optical tomography of photon Fock state superpositions
in connection with recent experimental achievements. The emphasis
is put on the fact that it suffices to represent the measured
tomogram as a main result of the experiment. We suggest a test for
checking the correctness of experimental data. Explicit
expressions for optical tomograms of Fock state superpositions are
given in terms of Hermite polynomials. Particular cases of vacuum
and low photon-number state superposition are considered as well
as influence of thermal noise on state purity is studied.
\end{abstract}


\bigskip

Last decades have seen a continuously growing interest to quantum
phenomena. The research has become interdisciplinary and covered
many fields from quantum optics and particle physics to quantum
information processing and foundations of quantum mechanics.
Preparation, evolution, and measurement of quantum states are
focused on in numerous theoretical and experimental
investigations. Also, a great effort was made in characterization
of quantum state properties such as nonlocality and entanglement.
Both explanation and prediction of experimental results resort to
a mathematical description of quantum states. Dealing with quantum
states of the electromagnetic field, widely used are the following
descriptions: the density operator $\hat{\rho}$ and several
quasidistribution functions like Wigner function. This is the
Wigner function that is the main goal of many scientific
laboratories interested in quantum state reconstruction. However,
the probabilistic nature of quantum mechanics does not allow
measuring Wigner function {\it directly} in the experiment. It can
only be obtained by subsequent processing of data. In view of this
fact, the question arises itself whether there exists such a
probability description of quantum states which is accessible for
a direct measuring. If such a description exists, then it can be
considered as an advantageous notion of quantum states.
Surprisingly, such a description does exist and it is an optical
tomogram~\cite{tombesi-manko,ibort}. This means that there is no
need to be aimed at obtaining Wigner function anymore.

Indeed, it was clarified recently~\cite{tombesi-manko,ibort} that
any quantum state can be identified with the optical tomogram.
From the mathematical point of view, the optical tomogram is
nothing else but Radon transform of the Wigner function. Crucial
point is that the Radon transform provides probability
distributions (see, e.g.~\cite{militello-manko}) and such a
distribution turns out to be measurable directly via the homodyne
detection of photon states. The experimental output of such
detection is exactly the optical tomogram. Originally, it was
used~\cite{raymer,lvovsky} as a technical tool to reconstruct the
Wigner function which is still usually identified with the photon
quantum state. On the other hand, the optical tomogram contains
complete information about all properties of a quantum state. In
view of this fact, it is pointed out that the accent in homodyne
detection experiments has to be put on the most accurate measuring
of the optical tomogram. According to association ``quantum state
$\Longleftrightarrow$ measurable tomogram", it would be enough to
present as a result of the experiment a plot of the tomogram.

Once optical tomogram is measured, there is a necessity to be
aware of output data being indeed a tomogram of quantum state. In
order to test correctness of the measured tomogram, one can
utilize peculiar properties of quantum states. For example, one
can check that the experimental results do not contradict to
tomographic entropic inequalities~\cite{DeNicola}, uncertainty
relations~\cite{mmsv} or purity constraints~\cite{porzio}. In
other words, one must check that the tomogram fulfils specific
requirements derived in~\cite{DeNicola,mmsv,porzio,mmanko}.
Apparently, these constraints are only necessary conditions but
their violation serves as a direct evidence of either experiment
or quantum theory being incorrect. In this paper, we will also
present an additional necessary condition on optical tomogram and
an estimation of the data accuracy.

The exciting progress of experimental techniques during last
several years gave rise to creation of paradigmatic photon states
like Fock state superpositions (FSS)
\cite{nature09,naturephotonics10,jila-93,haroche}. They are
nothing else but dramatic representatives of nonclassical
states~\cite{dodonov}. The above discussion on optical tomogram as
a primary notion of quantum states makes it reasonable to consider
the optical tomography of FSS. The aim of our paper is to fill a
gap between the experiment and the theoretical description of FSS.
Namely, we obtain an explicit expression for the optical tomogram
of FSS, analyze a highly simplified model of the influence of
thermal noise on optical tomograms and purity of FSS. We also
introduce a lower error bar on measured tomogram. The results of
this consideration altogether with the outstanding techniques like
those used in
experiments~\cite{nature09,naturephotonics10,jila-93,haroche,ourjoumtsev,zavatta,eichler}
could find application in further investigations for comparing
output data with predicted ones, estimating purity and thermal
noise presented, adjusting and calibrating the setup.

Optical tomogram of the state given by density operator
$\hat{\rho}$ reads

\begin{equation}
\label{optical-tomogram} w(X,\theta) = \int G(X,\theta;z,0)
G^{\ast}(X,\theta;z',0) \langle z | \hat{\rho} | z' \rangle {\rm
d}z {\rm d}z',
\end{equation}

\noindent where $G(q,t;q',0)$ is a conventional Green's function
of the harmonic oscillator with Hamiltonian $\hat{H} = (\hat{p}^2
+ \hat{q}^2)/2 = (\hat{a}^{\dag}\hat{a}+1/2)$. Hereafter we use
dimensionless units, namely, the Planck constant $\hbar=1$, the
Boltzmann constant $k_{\rm B} = 1$, etc. Formula
(\ref{optical-tomogram}) can be treated as a reconstruction of the
quantum state of wave packet from position probability
distributions measured during the packet's motion in the harmonic
oscillator potential~\cite{leonhardt,delcampo}. In case of optical
tomography, the timelike evolution of the electromagnetic field is
brought about by shifting the phase $\theta$ of the local
oscillator. The tomogram $w(X,\theta)$ is the marginal
distribution of the quadrature component $X$ of the electric field
strength, with $X$ being rotated by angle $\theta$ in the
quadrature phase space.

Let us consider the photon state of one-mode electromagnetic field
$|\psi\rangle=\sum_{n=0}^{N} c_n |n\rangle$, where $|n\rangle$ is
the photon Fock state, $N<\infty$. Density operator of such a pure
state is

\begin{equation}
\label{rho} \hat{\rho} = | \psi \rangle \langle \psi | =
\sum_{n=0}^{N} |c_n|^2 | n \rangle \langle n | + \sum_{n<k} \left(
c_n^{\ast} c_k | k \rangle \langle n | + {\rm h.c.} \right).
\end{equation}

\noindent The Green's function $G(q,t;q',0)$ determines an
evolution of Hamiltonian eigenstates $|n\rangle$ as follows

\begin{eqnarray}
\label{evolution} \int G(q,t;q',0) \langle q'| n \rangle {\rm d}q'
= \langle q| n \rangle e^{-i(n+1/2)t} =
\frac{1}{\sqrt{\sqrt{\pi}2^n n!}} H_{n}(q) e^{-q^2/2-i(n+1/2)t},
\end{eqnarray}

\noindent where $H_{n}(q)$ is the Hermite polynomial of degree
$n$. Substituting (\ref{rho}) and (\ref{evolution}) in
(\ref{optical-tomogram}), we readily obtain the optical tomogram
of Fock state superposition

\begin{eqnarray}
\label{optical-tomogram-FSS} w_{\rm FSS}(X,\theta) &=&
\frac{e^{-X^2}}{\sqrt{\pi}} \Bigg[ \sum_{n=0}^{N}
\frac{|c_n|^2}{2^n n!} H_n^2(X) + \sum_{n<k} \frac{\left(
c_n^{\ast} c_k e^{i(n-k)\theta}+ {\rm c.c.} \right)}{\sqrt{2^{n+k}
n!k!}} H_n(X) H_k(X) \Bigg]\nonumber\\
&=& \frac{e^{-X^2}}{\sqrt{\pi}} \Bigg[ \sum_{n=0}^{N}
\frac{|c_n|^2}{2^n n!} H_n^2(X) + \sum_{n<k} \frac{ |c_n| |c_k|
\cos ( (n-k)\theta - (\varphi_n-\varphi_k))}{\sqrt{2^{n+k-2}
n!k!}} H_n(X) H_k(X) \Bigg],
\end{eqnarray}

\noindent where we have extracted phases from coefficients
$c_n=|c_n|e^{i\varphi_n}$ and $c_k=|c_k|e^{i\varphi_k}$.

\begin{figure}
\begin{center}
\includegraphics{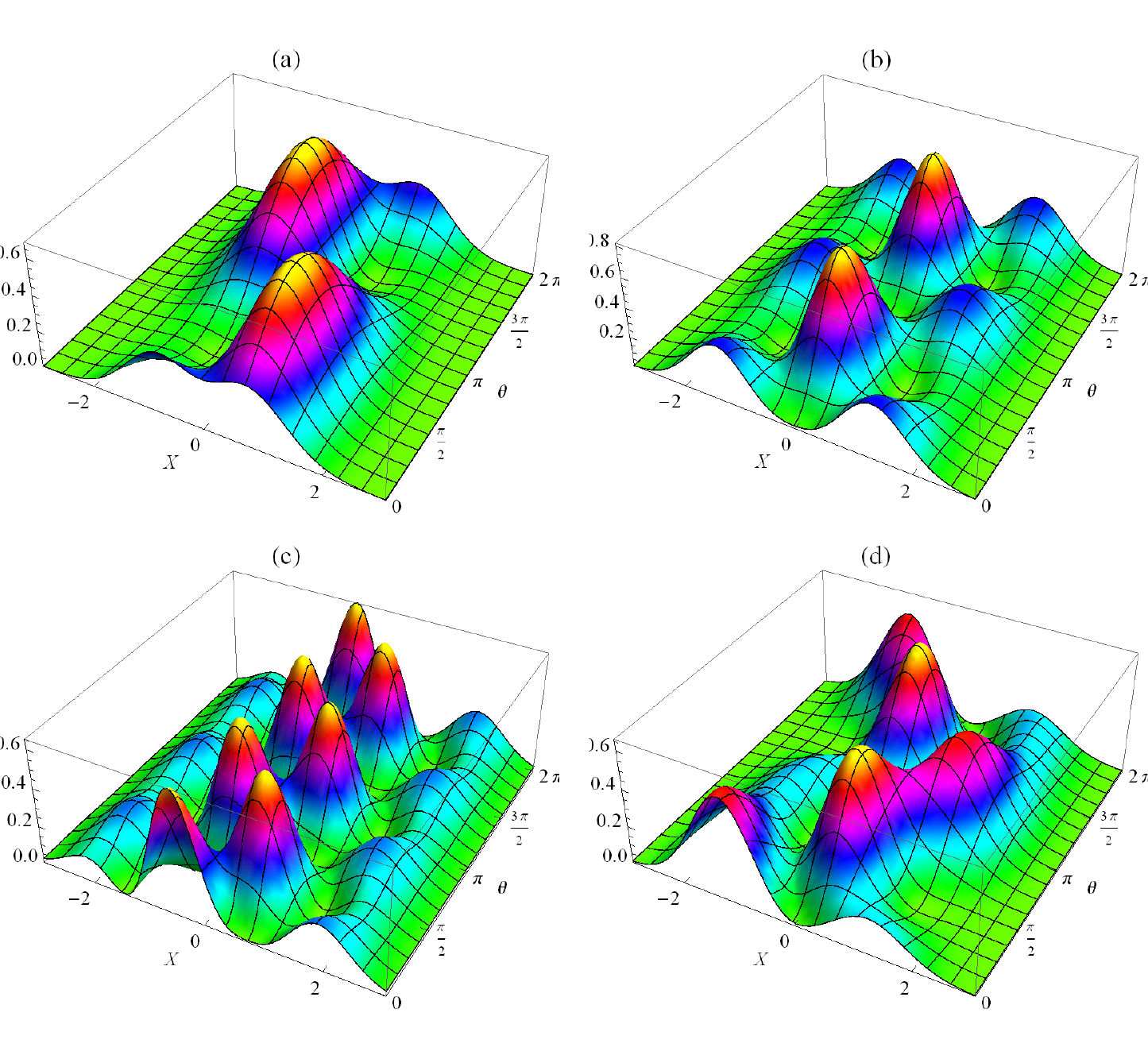}
\caption{\label{figure1} Optical tomograms $w(X,\theta)$ of Fock
state superpositions: $(|0\rangle+i|1\rangle)/\sqrt{2}$ (a),
$(|0\rangle+|2\rangle)/\sqrt{2}$ (b),
$(|0\rangle+|3\rangle)/\sqrt{2}$ (c), and
$(|0\rangle+e^{i2\pi/3}|1\rangle+|2\rangle)/\sqrt{3}$ (d).}
\end{center}
\end{figure}

Examples of FSS optical tomograms are illustrated in Fig.
\ref{figure1}. It is readily observed that the tomograms satisfy
the relation $w(X,\theta) = w(-X,\theta+\pi)$, which also follows
from analytical consideration. This simple constraint on
experimental data allows checking the accuracy of the measured
tomogram $w^{\rm meas}(X,\theta)$. Indeed, the difference $[w^{\rm
meas}(X,\theta) - w^{\rm meas}(-X,\theta+\pi)]$ must be zero for
an exactly measured tomogram. The deviation of this quantity from
the zero level for all the local oscillator phases, e.g.
$\sup_{X\in\mathbb{R},\theta\in[0,2\pi]} |w^{\rm meas}(X,\theta) -
w^{\rm meas}(-X,\theta+\pi)|$, is suggested to use as an indicator
of homodyne detection precision.

\begin{figure}
\begin{center}
\includegraphics{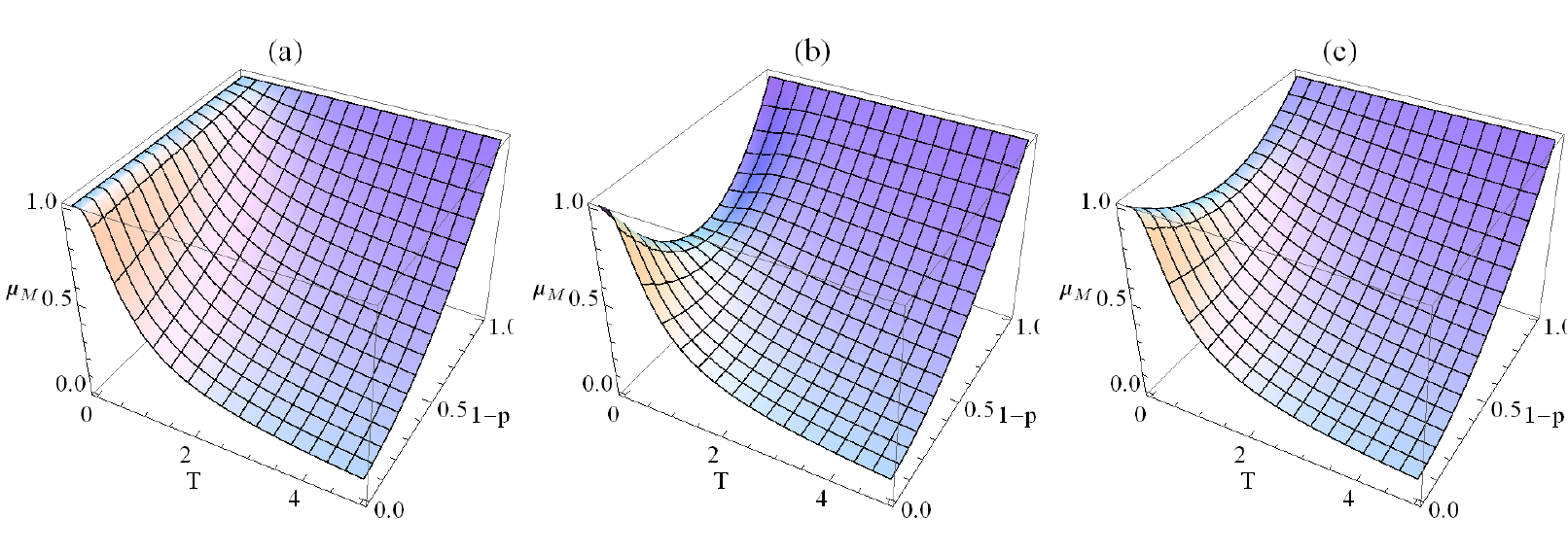}
\caption{\label{figure2} Purity $\mu_{\rm mix}$ of mixture of Fock
state superpositions with thermal noise vs. temperature $T$ and
mixture coefficient $p$: vacuum state $|0\rangle$ (a), one-photon
state $|1\rangle$ (b), and
$(|0\rangle+e^{i\varphi_1}|1\rangle)/\sqrt{2}$ (c).}
\end{center}
\end{figure}

It is worth mentioning that the Pauli problem can be
constructively solved for FSS~\cite{orlowski}. In other words, for
FSS it is sufficient to measure $w(X,\theta=0)$ and
$w(X,\theta=\pi/2)$ to find the initial state, up to the
well-known twofold ambiguity. This result is based on the fact
that we know a priori that the state of the system is pure and
comprises a finite number of Fock states. However, in real
experiments perfectly pure states are hardly achievable. For
instance, optical  tomography of microwave electromagnetic field
exploits linear amplifiers which unavoidably introduce a noise. We
will treat this noise as thermal one with the effective
temperature $T$. We find it reasonable to consider the influence
of such thermal noise on the purity of the desired FSS. Moreover,
the purity can be directly calculated via measured optical
tomogram~\cite{porzio} avoiding the density matrix formalism.
Comparison of obtained and theoretically predicted values of
purity can be used for testing the correctness of data,
calibrating and adjusting the apparatus.

For the sake of simplicity, we consider the following model. The
density operator of thermal noise has the form $\hat{\rho}_{\rm
th} = Z^{-1} e^{-\hat{H}/T}$, where $Z={\rm Tr}[e^{-\hat{H}/T}] =
\frac{1}{2}(\sinh\frac{1}{2T})^{-1}$ and $T$ is a temperature. The
associated optical tomogram reads $w_{\rm T}(X,\theta) =
(\sqrt{2\pi\sigma^2})^{-1/2}e^{-X^2/2\sigma^2}$, where $\sigma^2 =
\frac{1}{2}\coth\frac{1}{2T}$. Mixture of the superposition of
Fock states (\ref{rho}) and thermal noise is given by density
operator $\hat{\rho}_{\rm mix} = (1-p) \hat{\rho} + p
\hat{\rho}_{\rm th}$, where the parameter $0\le p \le 1$ shows how
much noise is added. The corresponding optical tomogram reads
$w_{\rm mix}(X,\theta) = (1-p)w_{\rm FSS}(X,\theta) + p w_{\rm
th}(X,\theta)$. Purity of this state is

\begin{eqnarray}
\mu_{\rm mix} &\equiv& {\rm Tr} \big[ \hat{\rho}_{\rm mix}^2 \big]
= (1-p)^2 + 2(1-p)p \langle \psi | \hat{\rho}_{\rm T} | \psi
\rangle + p^2
{\rm Tr} \big[ \hat{\rho}_{\rm th}^2 \big] \nonumber\\
&=& (1-p)^2 + 4(1-p)p \sinh\frac{1}{2T}\sum_{n=0}^{N} |c_n|^2
e^{-(n+\frac{1}{2})/T} + p^2 \tanh \frac{1}{2T}.
\end{eqnarray}

\noindent where we have taken into account the diagonal form of
thermal noise $\hat{\rho}_{\rm T}$ in the basis of Fock states.
Dependence of state purity on thermal noise is depicted for some
examples in Fig. \ref{figure2}. If $T\rightarrow \infty$, then
$\mu_{\rm mix} \approx (1-p)^2 + p^2/2T$. If $T\rightarrow 0$,
then the thermal noise reduces to vacuum mode and its effect on
purity depends on two factors: the noise strength $p$ and whether
the vacuum state was included in the initial FSS. If vacuum state
is absent in FSS, then $\mu_{\rm mix}(T=0)$ is the same for all
such FSS. It means that the parameter $p$ can be evaluated by
one-photon state. It is worth mentioning a non-monotonic
dependence of one-photon state purity on $T$ for a fixed $p<1/2$
because $\mu_{\rm mix}\approx (1-p)^2+p^2+2p(1-2p)e^{-1/T}$ if $T
\ll 1$ (see Fig. \ref{figure2}b). In the paper~\cite{militello},
the effect of purity oscillations is discussed. In present work we
note that the formula for purity in terms of optical tomograms can
also be used to describe such phenomena.

To conclude, we summarize the main results of the paper. We
studied the optical tomograms of Fock state superpositions in view
of several
experiments~\cite{nature09,naturephotonics10,jila-93,haroche}
devoted to measuring the homodyne quadrature distributions in some
nonclassical photon states~\cite{dodonov}. We point out that the
efforts of experiments should be focused on as precise measurement
of optical tomograms as possible. The optical tomogram provides
the complete information about a quantum state and the correctness
of measured tomogram can be checked by a series of constraints,
one of which is introduced in this paper. The other tests of the
precision of homodyne detecting photon states can be also
used~\cite{DeNicola,mmsv,porzio,mmanko}. We emphasize that there
is no need to convert the tomogram into Wigner function or other
quasi-probability functions. Being applied to Fock state
superpositions, the optical tomogram is obtained in explicit form
(\ref{optical-tomogram-FSS}). We have also analyzed the purity of
a Fock state superposition mixed with a thermal noise unavoidably
presented in experiments.

The authors appreciate the referees for useful remarks. The
authors thank the Russian Foundation for Basic Research for
partial support under Projects Nos. 09-02-00142 and 10-02-00312.
S.N.F. thanks the Russian Science Support Foundation for support
under Project ``Best postgraduates of the Russian Academy of
Sciences 2010" and the Ministry of Education and Science of the
Russian Federation for support under Projects Nos. 2.1.1/5909,
$\Pi$558, and 14.740.11.0497.

\end{document}